\documentclass[runningheads]{llncs}
\usepackage[T1]{fontenc}
\usepackage{dsfont}
\usepackage{amsmath,amssymb}
\usepackage{graphicx}
\usepackage{caption}
\usepackage{subcaption}
\usepackage{multirow}
\usepackage{makecell}
\usepackage{ragged2e}
\usepackage{xcolor}
\newcolumntype{x}[1]{>{\centering\let\newline\\\arraybackslash\hspace{0pt}}p{#1}}
\begin{document}
\title{SB-SSL: Slice-Based Self-Supervised Transformers for Knee Abnormality Classification from MRI}
\titlerunning{SB-SSL: Slice-Based Self-Supervised Transformers}
%
\author{Sara Atito\inst{1}\orcidID{0000-0002-7576-5791} \and
Syed Muhammad Anwar\inst{2}\orcidID{0000-0002-8179-3959} \and
Muhammad Awais \inst{1,3} \orcidID{0000-0002-1122-0709} \and Josef Kittler \inst{1} \orcidID{0000-0002-8110-9205}}

\institute{Centre for Vision, Speech and Signal Processing (CVSSP), University of Surrey, Guildford, United Kingdom \and Children's National Hospital, Washington, DC, USA \and
Surrey Institute for People-Centred AI}

\authorrunning{Atito et al.}

\maketitle              
\begin{abstract}
\vspace{-7mm}
The availability of large scale data with high quality ground truth labels is a challenge when developing supervised machine learning solutions for healthcare domain. Although, the amount of digital data in clinical workflows is increasing, 
most of this data is distributed on clinical sites and protected to ensure patient privacy. Radiological readings and dealing with large-scale clinical data puts a significant burden on the available resources, and this is where machine learning and artificial intelligence play a pivotal role. 
Magnetic Resonance Imaging (MRI) for musculoskeletal (MSK) diagnosis is one example where the scans have a wealth of information, but require a significant amount of time for reading and labeling. Self-supervised learning (SSL) can be a solution for handling the lack of availability of ground truth labels, but generally requires a large amount of training data during the pretraining stage. Herein, we propose a slice-based self-supervised deep learning framework (SB-SSL), a novel slice-based paradigm for classifying abnormality using knee MRI scans. 
We show that for a limited number of cases (<1000), our proposed framework is capable to identify anterior cruciate ligament tear with an accuracy of 89.17\% and an AUC of 0.954, outperforming state-of-the-art without usage of external data during pretraining. This demonstrates that our proposed framework is suited for SSL in the limited data regime.       

\keywords{Self-Supervised Learning  \and Group Masked Model Learning \and Masked Autoencoders \and Knee Abnormality \and Transformers \and MRI.}
\end{abstract}

\section{Introduction}
\label{sec:intro}
\vspace{-3mm}
Knee abnormality can arise from a variety of factors including aging, physical injury, and joint disease. MRI is the standard-of-care for diagnosis of knee abnormalities \cite{nacey2017magnetic}, where the image contains a wealth of information and the scanning protocols are safe from a clinical perspective. Knee MRI exams are among the most widely performed scans in MSK radiology \cite{irmakci2019deep}. MSK conditions arise from a variety of reasons (including sports injury and lifestyle choices) 
effecting adults and pediatrics. Both the amount of information within a knee MRI scan, and the number of such scans performed on a daily basis put a huge burden on the radiologist and the clinical workforce dealing with MSK related conditions 
and knee abnormalities 
In recent years, machine learning is the technology of choice in radiology for automated image analysis and abnormality identification \cite{anwar2018medical}. However, the clinical translation of this technology is facing challenges such as lack of adequate annotations and training data. In particular, manual segmentation and data labeling is a labor intensive and tedious task, which is also effected by inter-rater variability. The probability of error, accounting for the day-to-day workload on radiologists, is high and this is where machine learning can benefit the most by identifying the most critical cases needing immediate attention.

In contrast to Convolutional Neural Networks (CNNs), 
transformer-based deep learning models have shown to perform better due to an inherent design incorporating attention and parallel computing \cite{liu2021survey}. The success of transformer based networks in the field of natural language processing (NLP) is phenomenal and became the default choice in most recent NLP applications. The recent introduction of vision transformer  \cite{dosovitskiy2020image}, has resulted in the translation of some of this success to vision tasks. Training self-supervised vision transformers for medical applications could alleviate some of the problems associated with acquiring high quality ground truth labels and hence, accelerate the research in computer aided diagnosis. However, such networks require a large training data. Therefore, in Computer Vision (CV) problems, the default practice is to use a pretrained model on a large supervised data like ImageNet-1K, before fine tuning for a specific downstream task with limited data \cite{atito2021sit}. 

Recently, self-supervised pretraining of deep neural networks without using any labels has outperformed supervised pretraining in CV~\cite{atito2021mcssl,atito2022gmml}. This phenomenal shift in CV is less investigated in medical image analysis domain. We argue that recent SSL approaches are ideally suited for medical image analysis, since medical data are an order of magnitude smaller than natural images due to several reasons, including privacy concerns, expensive annotation, rarity of certain diseases, etc. 
Hence for medical applications, SSL can lead the way for a wider adoption of such techniques in domains where labels are not available or are difficult to acquire \cite{anwar2020semi}. Therefore, the purpose of this study is to investigate: 1) is ImageNet-1K pretraining needed for medical imaging? 2) can we perform self-supervised pretraining on a small medical data and outperform large scale out of distribution supervised pretraining? If successful this will form the basis for SSL for medical imaging in limited data regimes.
Towards this, we propose a slice-based self-supervised deep learning framework (SB-SSL) for abnormality classification using knee MRI, where our main contributions are:
\begin{itemize}
    \item We propose a novel slice based self-supervised transformer model (SB-SSL) for knee abnormality classification using magnetic resonance imaging data.
    \item The model is pretrained from scratch on limited data without labels and fine tuned for the downstream knee abnormality classification task with state-of-the-art performance.
    \item Our experimental results show that, when trained using the group masked model learning (GMML) paradigm, SSL can be successfully applied for medical image analysis with limited data/label.  
\end{itemize}

\section{Related Works}
\label{sec:related}

In \cite{bien2018deep}, a deep learning based method was presented for the detection of abnormalities in knee MRI. The publicly available MRNet data was presented, along with an AlexNet \cite{krizhevsky2012imagenet} based model for 
classifying abnormalities, meniscal tear, and anterior cruciate ligament (ACL) tear. This was among the first approaches where deep learning was applied to this task, and since then has been used in multiple studies to further improve the classification performance \cite{manna2022self,tsai2020knee,hung2022automatic,dunnhofer2021improving}.   

A CNN based self supervised training paradigm was developed, where solving the jigsaw puzzle was used as the pre-text task \cite{manna2022self}. In the downstream task, ACL tear was classified with an accuracy of $76.62\%$ and an area under the curve (AUC) of 0.848 using the sagittal plane. In \cite{tsai2020knee}, efficiently-layered network (ELNet) was proposed where the model reduced the number of parameters compared to AlexNet, and utilized individual slice views for classification of meniscus (coronal) and ACL (axial) tears. An accuracy of 0.904 with an AUC of 0.960 was achieved in detecting the ACL tear. This performance was improved by adding a feature pyramid network and pyramidal detail pooling to ELNet \cite{dunnhofer2021improving}. An AUC of 0.976 and an accuracy of 0.886 was achieved in ACL tear classification task. However, both these methods are based on supervised training. Meniscus tears were identified using a deep learning model and compared with manual evaluation \cite{hung2022automatic}. An accuracy of $95.8\%$ was achieved for an internal validation set, however the model was not evaluated on any of the publicly available data.

In general, it should be noted that for methods that report higher performance, training is based on the availability of ground truth labels. 
Whereas for self supervised training, which could alleviate this burden, the model performance drops. We propose, for the first time, a transformer based self-supervised framework for knee abnormality classification using MRI. Our innovative training paradigm use self-supervised training and shows that such a framework can be effectively used even when the size of training data is relatively small. 
\begin{figure*}[t]
    \centering
    \includegraphics[width=\linewidth]{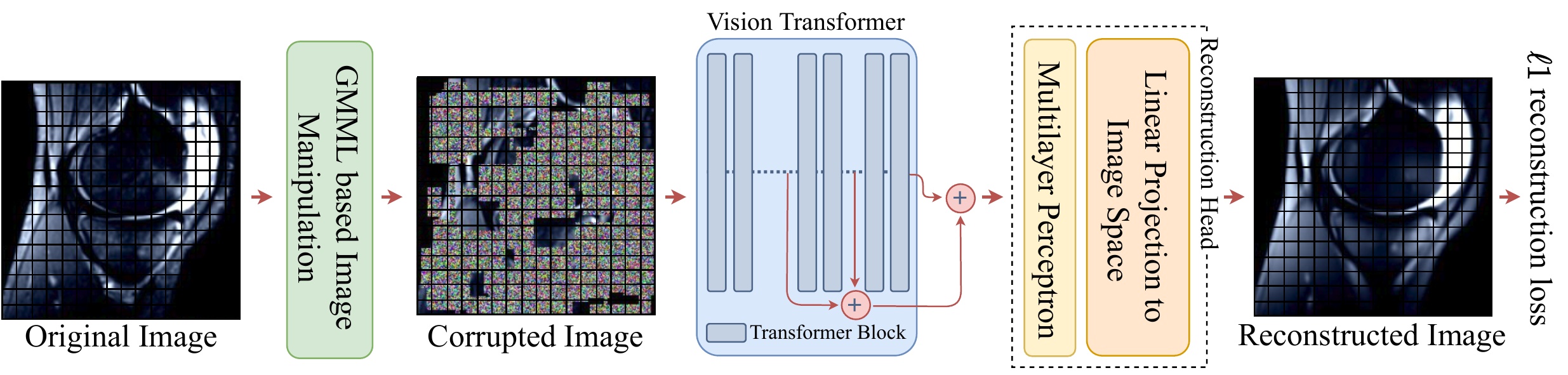}
    \caption{Proposed self-supervised learning approach.}
    \label{fig:GMMLtransformer}
\end{figure*}

\section{Methodology}
\label{sec:method}

In this work, we introduce a general slice-based self-supervised vision transformer for knee MRI medical records. The system diagram of the proposed approach is shown in Figure~\ref{fig:GMMLtransformer}. Transformers \cite{vaswani2017attention} have shown great success in various NLP and CV tasks \cite{devlin2018bert,radford2019language,brown2020language,atito2021sit,atito2021mcssl,atito2022gmml,zhou2022self,chen2022masked} and are the basis of our proposed framework. 

\subsection{Vision Transformer}
\label{sec:method_ViT}

Vision transformer \cite{dosovitskiy2020image} receives, as input, a feature map from the output of a convolutional block/layer with $K$ kernels of size $p \times p$ and stride $p \times p$. The convolutional block takes an input image $\mathbf{x} \in \mathcal{R}^{C \times H \times W}$ and converts it to feature maps of size $\sqrt{n} \times \sqrt{n} \times K$, where $C$, $H$, and $W$ are the number of channels, height, and width, of the input image, ($p \times p$) is the patch size, and $n$ is the number of patches, i.e., $n = \frac{H}{p} \times \frac{W}{p}$. Learnable position embeddings are added to the patch embeddings as an input to the transformer encoder to retain the relative spatial relation between the patches.

The transformer encoder consists of $L$ consecutive Multi-head Self-Attention (MSA) and Multi-Layer Perceptron (MLP) blocks. The MSA block is defined by $h$ self-attention heads, where each head outputs a sequence of size $n \times d$. 
The self attention mechanism is based on a trainable triplet (query, key, and value). Each query vector in $\mathbf{Q} \in \mathcal{R}^{n \times d}$  for a given head is matched against a set of key vectors $\mathbf{K} \in \mathcal{R}^{n \times d}$, scaled by the square root of $d$ to have more stable gradients as the dot product of $q$ and $k$ tend to grow large in magnitude, resulting in vanishing  gradients and a slowdown of learning. After applying softmax, the output is then multiplied by a set of values $\mathbf{V} \in \mathcal{R}^{n \times d}$. Thus, the output of the self-attention block is the weighted sum of $\mathbf{V}$ as shown in equation \ref{eq:SA}. The output sequences across heads are then concatenated into $n \times (d \times h)$, and projected by a linear layer to a $n \times K$ sequence. The MLP block consists of two point-wise convolution layers with GeLU~\cite{hendrycks2016gaussian} non-linearity.

\begin{equation}
    {\rm SelfAttention}(\mathbf{Q}, \mathbf{K}, \mathbf{V}) = {\rm Softmax}(\frac{\mathbf{Q}\mathbf{K}^T}{\sqrt{d}})\mathbf{V}.
    \label{eq:SA}
\end{equation}


\subsection{Self-supervised Pretraining}
\label{sec:method_ssl}

We leverage the strength of the transformers and train it as an autoencoder with a light decoder employing GMML \cite{atito2021sit,atito2022gmml}.  Starting with the vanilla transformer autoencoder, the model is pretrained as an autoencoder to reconstruct the input image, i.e., $D(E(\textbf{x})) = \textbf{x}$, where $\textbf{x}$ is the input image, $E$ is the encoder which is vision transformer in our case, and $D$ is a light reconstruction decoder. Due to the strength of transformers, it is expected that the model will perfectly reconstruct the input image after a few training epochs. Indeed, this is attributed to the fact that without a proper choice of constraints, autoencoders are capable of learning identity mapping, i.e., memorizing the input without learning any useful discriminative features. 

To promote the learning of context and learn better semantic representations of the input images from the transformer-based autoencoder, we apply several transformations to local patches of the image. The aim is to recover these masked local parts at the output of the light decoder. In doing so, especially with a high percentage of corruption (up to 70\%), the model implicitly learns the semantic concepts in the image and the underline structure of the data in order to be able to recover the image back. Image in-painting is a simple but effective pre-text task for self-supervision, which proceeds by training a network to predict arbitrary transformed regions based on the context. 

The objective of image reconstruction is to restore the original image from the corrupted image. For this task, we use the $\ell1$-loss between the reconstructed image and the original image in an end-to-end self-supervised trainable system as shown in equation \ref{eq:l1-pixel}. Although, $\ell2$-loss generally converges faster than $\ell1$-loss, it is prone to over-smooth the edges for image restoration \cite{zhao2016loss}. Therefore, $\ell1$-loss is more commonly used for image-to-image processing.
\begin{equation}
\label{eq:l1-pixel}
\mathcal{L}(\mathbf{W}) = \sum_k^b \left( \sum_i^H \sum_j^W \mathds{1}_{[\textbf{M}^k_{i,j} = 1]} | \mathbf{x}_{i,j}^k - \mathbf{\bar{x}}^k_{i,j} | \right),
\end{equation}
where $\mathbf{W}$ denotes the parameters to be learned during training, $b$ is the batch size, $\mathbf{M}$ is a binary mask with 1 indicating the manipulated pixels, and $\mathbf{\bar{x}}$ is the reconstructed image. To further improve the performance of the autoencoder, we introduced skip connections from several intermediate transformer blocks to the decoder. These additional connections can directly send the feature maps from the earlier layers of the transformers to the decoder which helps to use fine-grained details learned in the early layers to construct the image. Besides, skip connections in general make the loss landscape smoother which leads to faster convergence. 
Further, the reconstructed image $\mathbf{\bar{x}}$ is obtained by averaging the output features from the intermediate blocks from the transformer encoder ($E(.)$) and feeding the output to a light decoder ($D(.)$) represented mathematically as $\mathbf{\bar{x}} = D \left(\textstyle \sum_{i \in \mathcal{B}} E_i(\mathbf{\hat{x}}) \right)$, where $E_i(.)$ is the output features from block $i$ and $\mathcal{B}$ is a pre-defined index set of transformer blocks that are included in the decoding process. Herein, we set $\mathcal{B}$ to $\{6, 8, 10, 12\}$.

As for the decoder, unlike CNN-based autoencoders which require expensive decoders consisting of convolutional and transposed convolution layers, the decoder in the transformer autoencoder can be implemented using a light decoder design. Specifically, our decoder consisted of two point-wise convolutional layers with GeLU non-linearity and a transposed convolutional layer to return back to the image space. Since the backbone, i.e., vision transformer, and the light decoder are isotropic, some of the transformer blocks may act as decoder and hence, heavy and computationally expensive type of decoders are not required.

\section{Experimental Results}

To demonstrate the effectiveness of our proposed self-supervised vision transformer on medical images, we employed the MRNet dataset \cite{bien2018deep}. The dataset consists of 1,370 knee MRI records, split into a training set of 1,130 records of 1,088 patients and a validation set of 120 records of 111 patients. Each MRI is labeled according to the presence/absence of meniscus tear, ACL tear, or any other abnormality in the knee. In this work, we tackled the ACL tear identification problem using the Sagittal plane. The dataset is highly imbalanced with only 208 MRIs representing ACL tear.

\subsection{Implementation Details}
In our work, we employed the ViT Small (ViT-S) variant of the transformer \cite{touvron2020training} with $256\times256$ input image size. For optimization of the transformer parameters during self-supervised pre-training, we used the Adam optimizer~\cite{Loshchilov2017FixingWD} with a momentum of $0.9$. The weight decay follows a cosine schedule~\cite{loshchilov2016sgdr} from $0.04$ to $0.4$, and a base learning rate of $5e^{-4}$. All models were pre-trained employing $4$ Nvidia Tesla V100 32GB GPU cards with $64$ batch size per GPU. 

Simple data augmentation techniques were applied like random cropping, random horizontal flipping, random Gaussian blurring, and random adjusting of the sharpness, contrast, saturation, and the hue of the image. The augmented image was further corrupted by randomly replacing patches from the image with zeros, with a replacement rate of up to 70\% of the image pixels. 

For fine-tuning, we drop the light decoder and fine tune the pre-trained model by passing the volume, slice by slice, to the transformer encoder. The outputs of the class tokens corresponding to each slice are then concatenated to obtain $y \in \mathcal{R}^{f \times K}$, where $f$ is the number of slices. After that, the features $y$ are fed to a fully connected layer with $K$ nodes followed by GeLU non-linearity, followed by a linear layer with $2$ nodes corresponding to the presence/absence of the ACL tear. As the dataset is highly imbalanced, we used oversampling on the training set to balance the dataset. Specifically, we over-sample the minority class, i.e., presence of ACL tear, to match the number of the majority class. Finally, we applied the same optimization parameters and data augmentations used for the self-supervised training.

Further, we employed ensemble learning \cite{kittler1998combining}. Generally, neural networks have high variance due to the stochastic training approach that makes them sensitive to the nature of the training data. The models may find a different set of weights each time they are trained, which in turn may produce different predictions. To mitigate this issue, for each experiment, we trained $5$ models with different weight initialization and combined the predictions from these models. Not only this approach reduced the variance of the predictions, but also resulted in predictions that were better than any single model. 

\subsection{Results}
\label{sec:res}
\begin{figure}[h]
\centering
\begin{minipage}{0.6\textwidth}
\centering
\captionof{table}{Comparison with SOTA on ACL tears classification employing sagittal plane. }
\label{tbl:res}
\resizebox{\textwidth}{!}{
\begin{tabular}{p{2.5cm}x{1.5cm}x{1.5cm}x{2cm}x{2cm}}
\hline
\multicolumn{1}{c}{\multirow{3}{*}{Method}} &
\multicolumn{1}{c}{\multirow{3}{*}{Backbone}} &
\multicolumn{1}{c}{\multirow{3}{*}{\# params}} &
\multicolumn{2}{c}{ACL Tear (Sagittal plane)}        \\ \cline{4-5} 
&&\multicolumn{1}{c}{} & Accuracy& \multirow{2}{*}{AUC}\\ 
&&&(\%)&\\\hline
\multicolumn{5}{l}{\textit{\color{gray}{{Training using only the given dataset}}}} \\
Random Init       & CNN   &  77M & 71.67 & 0.754\\
Random Init       & ViT-S & 21M & 70.00 & 0.721\\ \cline{4-5} 
\cite{manna2022self}& CNN   &  77M & 76.62 & 0.848\\ 
\cite{manna2022self} + noise& CNN   &  77M & 75.83 & 0.817\\ \cline{4-5} 
SB-SSL (Ours)              & ViT-T & 5M  &  85.83 & 0.952\\
SB-SSL (Ours)               & ViT-S & 21M &  88.33 & 0.954\\
SB-SSL (Ours)               & ViT-B & 86M &  89.17 & 0.954\\\hline
\multicolumn{5}{l}{\textit{\color{gray}{{Transfer learning from ImageNet-1K dataset}}}} \\
MRNet \cite{bien2018deep}     & AlexNet & 61M & 86.63 & 0.963\\\hline
\end{tabular}}
\end{minipage}
\hspace{0.01\textwidth}
\begin{minipage}{0.37\textwidth}
\centering
\includegraphics[width=\textwidth]{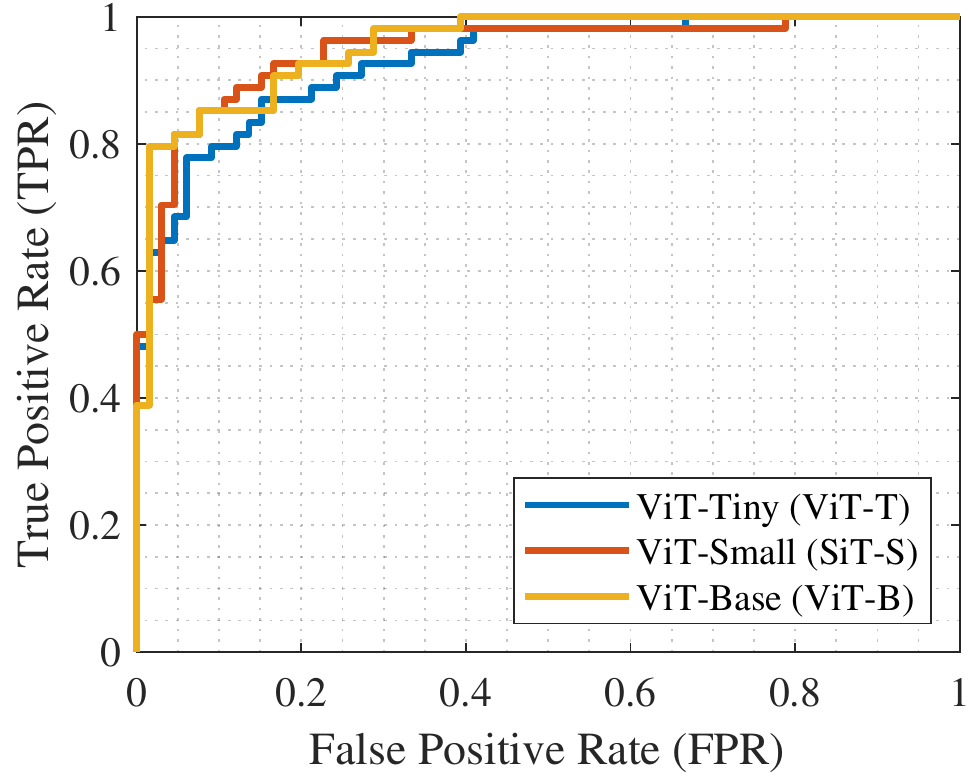}
\vspace{-0.6cm}
\caption{ROC curves of the classification of ACL tears employing different vision transformer architectures.}
\label{fig:ROC}
\end{minipage}
\end{figure}

It is well known that transformers are data-hungry which make them hard to train, mostly, due to the lack of the typical inductive bias of convolution operations.
Consequently, the common protocol for self-supervised learning with transformers is to pretrain the model on a large scale dataset, such as ImageNet or even larger datasets.  
We compare our proposed approach with the state-of-the-art SSL methods when the pretraining and the fine-tuning are performed only on the MRNet dataset. Table \ref{tbl:res} shows that our method outperforms the state-of-the-art with a large margin with an improvement of 12.6\% top-1 validation accuracy on the ACL tears classification task employing the sagittal plane. Most importantly, without using any external data, our proposed approach outperforms the competitors that are pre-trained with ImageNet-1K marking a milestone for the medical domain. The receiver operating characteristic (ROC) curve for three transformer variants, ViT-Tiny, ViT-Small, and ViT-Base are shown in Figure \ref{fig:ROC}, where ViT-T performs the best.

\vspace{-5mm}

\subsection{Ablation Studies}
\label{sec:Ablation}
In this section, we investigate the effect of different recipes of the proposed approach, such as the effect of longer pretraining, the size of the model, and the type of image corruption during the pretraining stage. Further, we show the interpretability of the system by visualizing the attention of the trained models.

\begin{figure}[!h]
\centering
\begin{subfigure}{0.45\linewidth}
 \includegraphics[width=\linewidth]{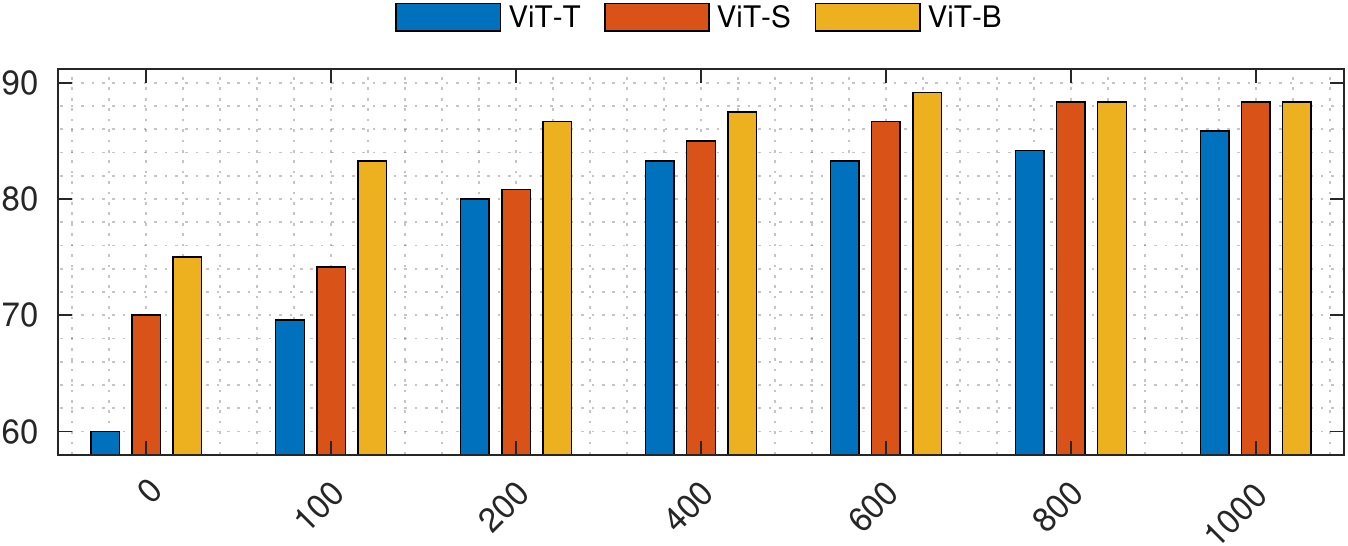}   
 \caption{Accuracy}
\end{subfigure}
\begin{subfigure}{0.45\linewidth}
\includegraphics[width=\linewidth]{Figures/ablation/Ablation_acc.pdf}
 \caption{AUC}
\end{subfigure}
\caption{Top-1 validation accuracies and AUC of the MRNet validation set across different vision transformer architectures. The x-axis represents number of epochs used for pretraining. }
\label{fig:longertraining}
\end{figure}

\noindent
\textbf{Effects of Longer Pretraining and Model Size.}
In Figure \ref{fig:longertraining}, we show the performance of the proposed approach when pretrained for longer duration across different vision transformer architectures. The x-axis represents the number of self-supervised pretraining epochs, with zero indicating that the model was not pretrained, i.e., training from scratch.
From the reported results, it is evident that the training from random initialization has produced a lower accuracy as the amount of data available is insufficient to train the transformer. The results significantly improved when the models were pretrained without any external data by 25.8\%, 18.3\%, and 13.3\% employing ViT-T, ViT-S, and ViT-B, respectively, compared to training from scratch. Another observation is that pre-training the self-supervised for longer and employing bigger transformer architectures contribute positively to the performance of the proposed approach.

\begin{figure*}[t]
    \centering

    \begin{subfigure}[t]{0.32\textwidth}
        \centering
        \includegraphics[width=\linewidth]{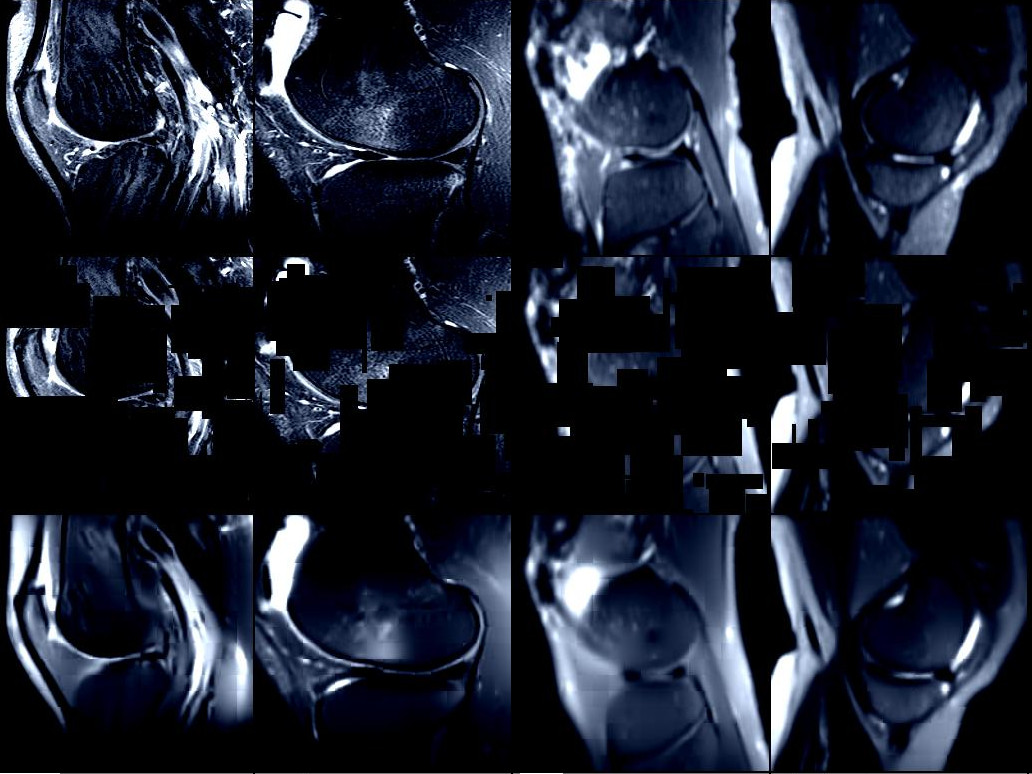}
        \caption{Zeros}
    \end{subfigure}%
    \hspace{0.02cm}
    \begin{subfigure}[t]{0.32\textwidth}
        \centering
        \includegraphics[width=\linewidth]{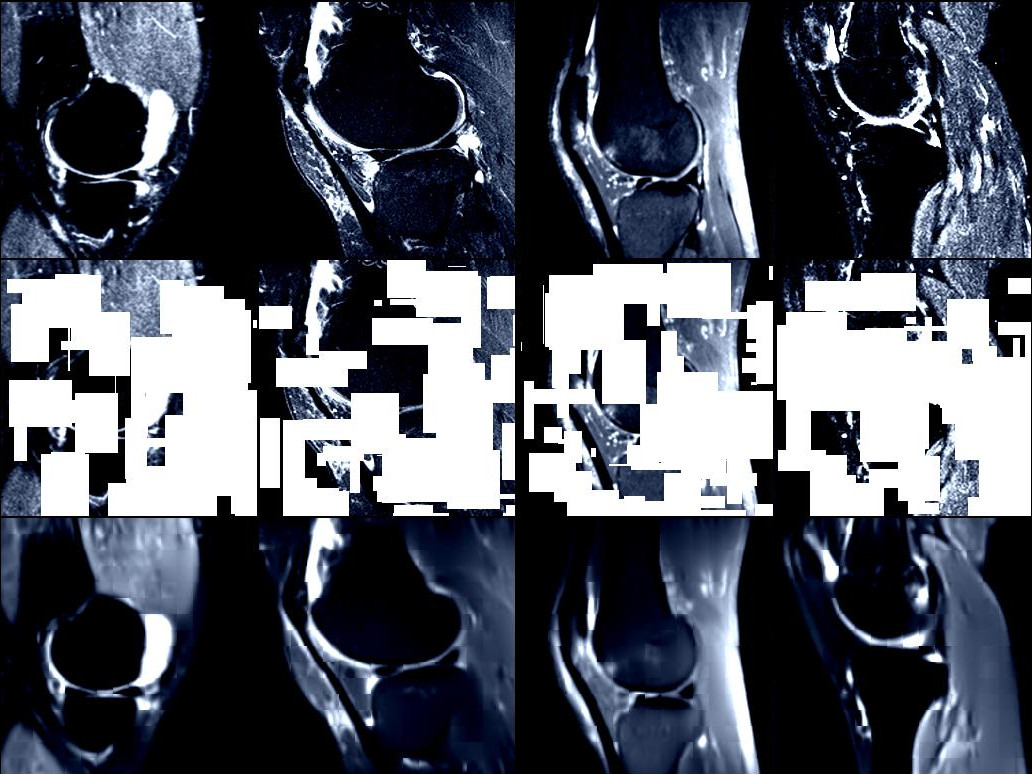}
        \caption{Ones}
    \end{subfigure}%
    \hspace{0.02cm}
    \begin{subfigure}[t]{0.32\textwidth}
        \centering
        \includegraphics[width=\linewidth]{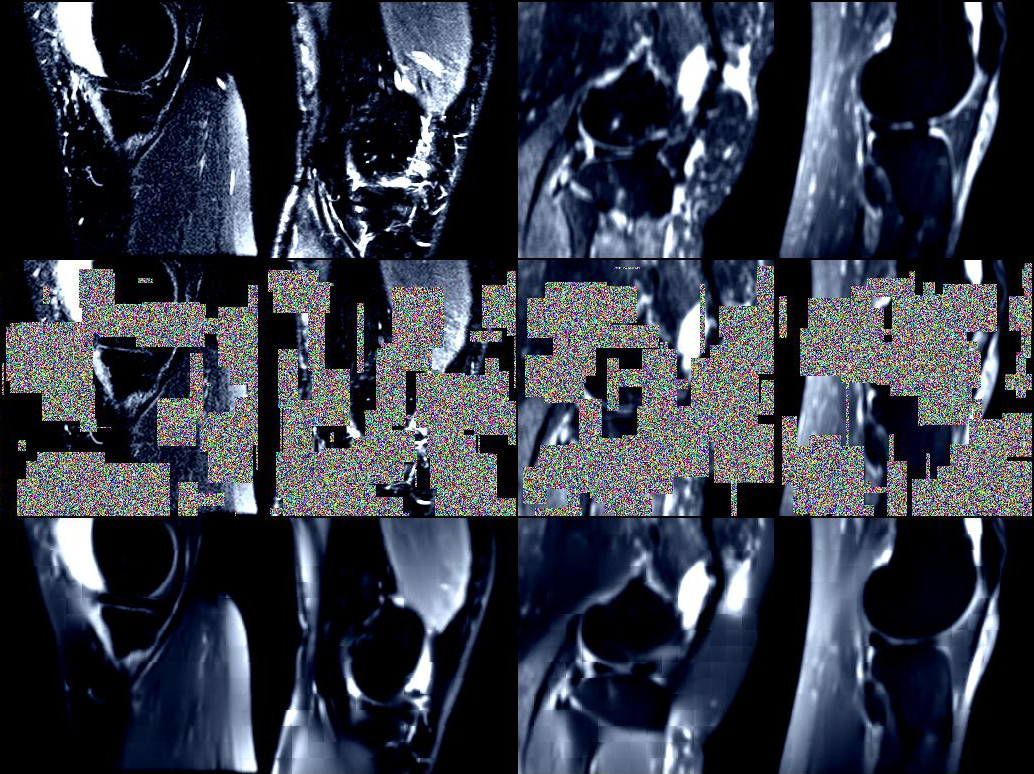}
        \caption{Noise}
    \end{subfigure}%
    
    \caption{Samples of different types of corruption. The rows represent the original images, corrupted images, and the reconstructed images after the pretraining stage, respectively.}
    \label{fig:drop_type}
\end{figure*}

\noindent \textbf{
The Effects of Different Types of Corruption:}
We first investigate the effect of training a vanilla transformer autoencoder, where the model is pretrained as an autoencoder to reconstruct the input image. As expected, after finetuning, the performance was  similar to the performance of the model trained from scratch. Following, we investigate the effect of applying different types of image inpainting including: random masking by replacing a group of connected patches from the image with zeros, ones, or noise. Samples of the different types of corruption are shown in Figure \ref{fig:drop_type} along with the reconstructed images after the pretraining stage. The performance when pretraining the models with different types of corruption is on par, with noise being marginally better than others. 

\begin{figure}[!h]
\centering
\includegraphics[width=0.092\linewidth]{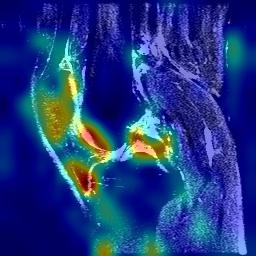}
\includegraphics[width=0.092\linewidth]{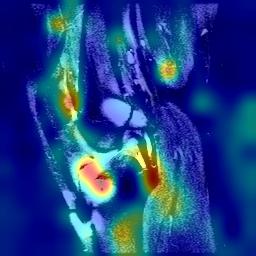}
\includegraphics[width=0.092\linewidth]{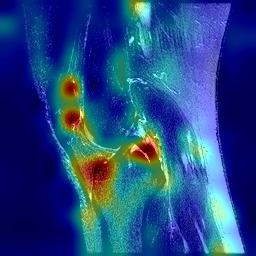}
\includegraphics[width=0.092\linewidth]{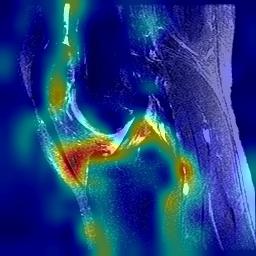}
\includegraphics[width=0.092\linewidth]{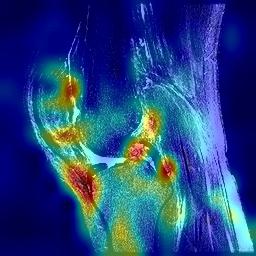}
\includegraphics[width=0.092\linewidth]{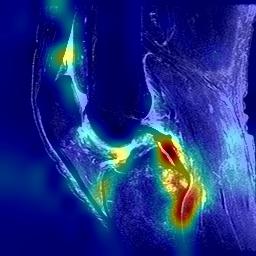}
\includegraphics[width=0.092\linewidth]{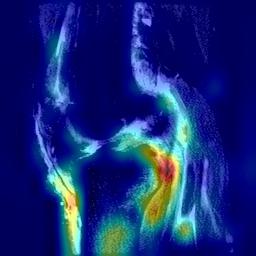}
\includegraphics[width=0.092\linewidth]{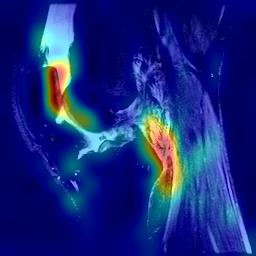}
\includegraphics[width=0.092\linewidth]{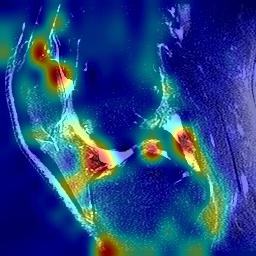}
\includegraphics[width=0.092\linewidth]{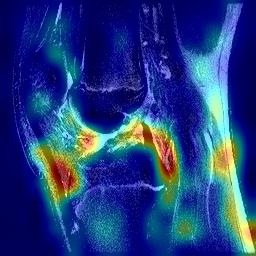}
\caption{Self-attention visualizations from the ViT-S model finetuned on the ACL tears task employing the sagittal plane.}
\label{fig:attn_vis}
\end{figure}

\noindent
\textbf{Attention Visualization.} To verify that the model is learning pertinent features, in Figure \ref{fig:attn_vis}, we provide visualizations of the self-attention corresponding to the class token of the $10^\text{th}$ layer of the vision transformer. To generate the attention for an image, we compute the normalized average over the self-attention heads to obtain a $16\times16$ tokens. The tokens are then mapped to a color scheme, up-sampled to $256\times256$ pixels, and overlaid with the original input image. For visualization, we selected the mid slice of randomly selected MRI volumes from the MRNet validation set. We observe that the attention is clearly focusing on the area of interest, corresponding to the main part of the MRI slice on which the detection of ACL tears is performed.

\section{Conclusion}
We proposed a novel framework SB-SSL, pre-trained in a self-supervised manner for knee abnormality classification. We established a new benchmark in SSL for MRI data, where pretraining on a large supervised data was not required. The state-of-the-art performance, with an accuracy of 89.17\% in ACL tear classification, shows that our proposed method can be employed in MR image classification even when the data are limited and ground truth labels are not available.

%
%
%
\bibliographystyle{splncs04}
\bibliography{biblo}

\begin{thebibliography}{10}
\providecommand{\url}[1]{\texttt{#1}}
\providecommand{\urlprefix}{URL }
\providecommand{\doi}[1]{https://doi.org/#1}

\bibitem{anwar2020semi}
Anwar, S.M., Irmakci, I., Torigian, D.A., Jambawalikar, S., Papadakis, G.Z.,
  Akgun, C., Ellermann, J., Akcakaya, M., Bagci, U.: Semi-supervised deep
  learning for multi-tissue segmentation from multi-contrast mri. Journal of
  Signal Processing Systems pp. 1--14 (2020)

\bibitem{anwar2018medical}
Anwar, S.M., Majid, M., Qayyum, A., Awais, M., Alnowami, M., Khan, M.K.:
  Medical image analysis using convolutional neural networks: a review. Journal
  of medical systems  \textbf{42}(11),  1--13 (2018)

\bibitem{atito2021mcssl}
Atito, S., Awais, M., Farooq, A., Feng, Z., Kittler, J.: Mc-ssl0. 0: Towards
  multi-concept self-supervised learning. arXiv preprint arXiv:2111.15340
  (2021)

\bibitem{atito2021sit}
Atito, S., Awais, M., Kittler, J.: Sit: Self-supervised vision transformer.
  arXiv preprint arXiv:2104.03602  (2021)

\bibitem{atito2022gmml}
Atito, S., Awais, M., Kittler, J.: Gmml is all you need. arXiv preprint
  arXiv:2205.14986  (2022)

\bibitem{bien2018deep}
Bien, N., Rajpurkar, P., Ball, R.L., Irvin, J., Park, A., Jones, E., Bereket,
  M., Patel, B.N., Yeom, K.W., Shpanskaya, K., et~al.: Deep-learning-assisted
  diagnosis for knee magnetic resonance imaging: development and retrospective
  validation of mrnet. PLoS medicine  \textbf{15}(11),  e1002699 (2018)

\bibitem{brown2020language}
Brown, T.B., Mann, B., Ryder, N., Subbiah, M., Kaplan, J., Dhariwal, P.,
  Neelakantan, A., Shyam, P., Sastry, G., Askell, A., et~al.: Language models
  are few-shot learners. arXiv preprint arXiv:2005.14165  (2020)

\bibitem{chen2022masked}
Chen, Z., Agarwal, D., Aggarwal, K., Safta, W., Balan, M.M., Sethuraman, V.,
  Brown, K.: Masked image modeling advances 3d medical image analysis. arXiv
  preprint arXiv:2204.11716  (2022)

\bibitem{devlin2018bert}
Devlin, J., Chang, M.W., Lee, K., Toutanova, K.: Bert: Pre-training of deep
  bidirectional transformers for language understanding. arXiv preprint
  arXiv:1810.04805  (2018)

\bibitem{dosovitskiy2020image}
Dosovitskiy, A., Beyer, L., Kolesnikov, A., Weissenborn, D., Zhai, X.,
  Unterthiner, T., Dehghani, M., Minderer, M., Heigold, G., Gelly, S., et~al.:
  An image is worth 16x16 words: Transformers for image recognition at scale.
  arXiv preprint arXiv:2010.11929  (2020)

\bibitem{dunnhofer2021improving}
Dunnhofer, M., Martinel, N., Micheloni, C.: Improving mri-based knee disorder
  diagnosis with pyramidal feature details. In: Medical Imaging with Deep
  Learning. pp. 131--147. PMLR (2021)

\bibitem{hendrycks2016gaussian}
Hendrycks, D., Gimpel, K.: Gaussian error linear units (gelus). arXiv preprint
  arXiv:1606.08415  (2016)

\bibitem{hung2022automatic}
Hung, T.N.K., Vy, V.P.T., Tri, N.M., Hoang, L.N., Tuan, L.V., Ho, Q.T., Le,
  N.Q.K., Kang, J.H.: Automatic detection of meniscus tears using backbone
  convolutional neural networks on knee mri. Journal of Magnetic Resonance
  Imaging  (2022)

\bibitem{irmakci2019deep}
Irmakci, I., Anwar, S.M., Torigian, D.A., Bagci, U.: Deep learning for
  musculoskeletal image analysis. In: 2019 53rd Asilomar Conference on Signals,
  Systems, and Computers. pp. 1481--1485. IEEE (2019)

\bibitem{kittler1998combining}
Kittler, J., Hatef, M., Duin, R.P., Matas, J.: On combining classifiers. IEEE
  transactions on pattern analysis and machine intelligence  \textbf{20}(3),
  226--239 (1998)

\bibitem{krizhevsky2012imagenet}
Krizhevsky, A., Sutskever, I., Hinton, G.E.: Imagenet classification with deep
  convolutional neural networks. Advances in neural information processing
  systems  \textbf{25} (2012)

\bibitem{liu2021survey}
Liu, Y., Zhang, Y., Wang, Y., Hou, F., Yuan, J., Tian, J., Zhang, Y., Shi, Z.,
  Fan, J., He, Z.: A survey of visual transformers. arXiv preprint
  arXiv:2111.06091  (2021)

\bibitem{loshchilov2016sgdr}
Loshchilov, I., Hutter, F.: Sgdr: Stochastic gradient descent with warm
  restarts. arXiv preprint arXiv:1608.03983  (2016)

\bibitem{Loshchilov2017FixingWD}
Loshchilov, I., Hutter, F.: Fixing weight decay regularization in adam. ArXiv
  \textbf{abs/1711.05101} (2017)

\bibitem{manna2022self}
Manna, S., Bhattacharya, S., Pal, U.: Self-supervised representation learning
  for detection of acl tear injury in knee mr videos. Pattern Recognition
  Letters  \textbf{154},  37--43 (2022)

\bibitem{nacey2017magnetic}
Nacey, N.C., Geeslin, M.G., Miller, G.W., Pierce, J.L.: Magnetic resonance
  imaging of the knee: an overview and update of conventional and state of the
  art imaging. Journal of Magnetic Resonance Imaging  \textbf{45}(5),
  1257--1275 (2017)

\bibitem{radford2019language}
Radford, A., Wu, J., Child, R., Luan, D., Amodei, D., Sutskever, I.: Language
  models are unsupervised multitask learners. OpenAI blog  \textbf{1}(8), ~9
  (2019)

\bibitem{touvron2020training}
Touvron, H., Cord, M., Douze, M., Massa, F., Sablayrolles, A., J{\'e}gou, H.:
  Training data-efficient image transformers \& distillation through attention.
  arXiv preprint arXiv:2012.12877  (2020)

\bibitem{tsai2020knee}
Tsai, C.H., Kiryati, N., Konen, E., Eshed, I., Mayer, A.: Knee injury detection
  using mri with efficiently-layered network (elnet). In: Medical Imaging with
  Deep Learning. pp. 784--794. PMLR (2020)

\bibitem{vaswani2017attention}
Vaswani, A., Shazeer, N., Parmar, N., Uszkoreit, J., Jones, L., Gomez, A.N.,
  Kaiser, L., Polosukhin, I.: Attention is all you need. arXiv preprint
  arXiv:1706.03762  (2017)

\bibitem{zhao2016loss}
Zhao, H., Gallo, O., Frosio, I., Kautz, J.: Loss functions for image
  restoration with neural networks. IEEE Transactions on computational imaging
  \textbf{3}(1),  47--57 (2016)

\bibitem{zhou2022self}
Zhou, L., Liu, H., Bae, J., He, J., Samaras, D., Prasanna, P.: Self
  pre-training with masked autoencoders for medical image analysis. arXiv
  preprint arXiv:2203.05573  (2022)

\end{thebibliography}
%

\end{document}